\documentclass[twocolumn,showpacs,preprintnumbers,amsmath,amssymb]{revtex4-1}
\usepackage{epsfig}
\usepackage{graphicx}
\usepackage{epsfig}
\usepackage{color}
\usepackage{rotating}
\usepackage{amssymb}
\usepackage{amsmath}
\usepackage{graphicx}

\begin{document}

\title{Global estimates of errors in quantum computation by the Feynman-Vernon formalism}

\author{Erik Aurell}
\email{eaurell@kth.se}
\affiliation{
Dept. of Computational Biology and Center for Quantum Materials, 
KTH -- Royal Institute of Technology,  AlbaNova University Center, SE-106 91~Stockholm, Sweden\\
Depts. Applied Physics and Computer Science, Aalto University, 
P.O. Box 11100, FI-00076 AALTO, Finland\\
Institute of Theoretical Physics, Chinese Academy of Sciences, 
P.O. Box 2735, Beijing 100190, China
}

\begin{abstract}
The operation of a
quantum computer is considered as a general quantum operation on a mixed state
on many qubits followed by a measurement. 
The general quantum operation is further 
represented as a Feynman-Vernon double path integral over the
histories of the qubits and of an environment, and afterward tracing out
the environment. The qubit histories are taken to be
paths on the two-sphere $S^2$ as in Klauder's coherent-state path integral
of spin, and the environment is assumed to consist of harmonic oscillators
initially in thermal equilibrium, and linearly coupled to to qubit operators $\hat{S}_z$.
The environment can then be integrated out to give a Feynman-Vernon
influence action coupling the forward and backward histories of the qubits.
This representation allows to derive in a simple way estimates that the total error of operation
of a quantum computer without error correction scales linearly 
with the number of qubits and the time of operation.
It also allows to discuss Kitaev's toric code interacting with an environment in the same manner.
\end{abstract}

\maketitle

\section{Introduction} 
\label{sec:introduction}
Quantum computers are physical devices that manipulate quantum states to execute
information-processing tasks~\cite{NielsenChuang2000,Mermin2007}. 
To build a general-purpose quantum computer is a difficult experimental challenge
where several different realizations have been proposed since the 
1990ies~\cite{Gershenfeld350,Ladd2010}.
While several large commercial initiatives to reach 
on the order of 50 qubits have been widely reported on recently~\cite{Courtland2017,Savage2017},
the current public state-of-the-art is that around ten qubits can be manipulated in the
lab in a manner approaching to what would be required for a general-purpose quantum 
computer~\cite{Kelly15}. 
For more restricted computational tasks a device using about 1000~qubits
has been reported to lead to important speed-up over classical algorithms~\cite{Denchev2016}.

Quantum computation would, if successful, upend the characterization 
of what is possible and not possible in classical information processing.
That is its main promise, but also a main difficulty
since a large quantum computer must have a substantial number of degrees of freedom, and 
large physical systems have a strong tendency to turn classical~\cite{HarocheRaimond1996}.
The underlying mechanism of this quantum-to-classical transition is the decoherence
of the quantum state by interactions between the quantum computer and the
rest of the world~\cite{Zurek2003}. 

There has been a long-running polemic against the possibility of quantum computing 
going back at least to~\cite{Alickietal2002}.
This paper is primarily concerned with the critique recently put forward in~\cite{Kalai-2016},
to be discussed again in Sections~\ref{sec:relation-to-literature}
and~\ref{sec:pessimistic-hypothesis} below.
The gist of these arguments is that the standard error models considered in
the quantum computing literature are too restrictive, and that quantum computing
will for this reason not be possible in the real world when errors may be
correlated in time and space.
Indeed, errors in quantum computation are often discussed as mis-application of operators,
which indicates a kind of classical uncertainty in the external potential controlling the quantum 
system, and not errors induced by
coupling to another (unobserved) quantum system.

The goal of this paper is to consider this problem 
from a global point of view
by investigating the errors caused by coupling 
a system of spins to a thermal
bath of bosonic degrees of freedom. Such errors can be correlated over
arbitrary distances at low enough temperature, but in a specific way 
determined by the physical interaction. 
The analysis is carried out by combining a coherent-state path integral representation
of the dynamics of spin systems with the Feynman-Vernon method to integrate 
out the bath. The effect of the bath is then described by an
influence functional coupling forward and backward coherent-state path integrals.
This influence functional can be estimated simply when the system-bath interaction 
is weak. 

The main result is that the errors usually considered are not an idealization but
rather a worst-case scenario. It is correct that these standard error models mostly 
disregard correlations in time and space, but a physical heat bath, which could generate such correlations,
is a comparatively simple system, and not an adversary in the sense of complexity theory.
For the paradigmatic example of the Kitaev toric code it is further showed that 
coupling to a bath has effects exponentially small in the size of the 
lattice. For such a system only the errors normally considered therefore
need to be corrected. 

The paper is organized as follows: in Section~\ref{sec:relation-to-literature}
I position the problem in relation to the quantum computational literature
and in Section~\ref{sec:pessimistic-hypothesis}
I state ``pessimistic hypothesis'' of Kalai in a version suitable to
be analyzed by the methods used herein.
In Section~\ref{sec:quantum-noise} I review quantum noise, and in
Section~\ref{sec:open-quantum-model} I introduce
the Feynman-Vernon double path integral as a model of
quantum operations on $n$ qubits which also interact linearly
with a bath of harmonic oscillators.
The system qubit
histories are formulated using Klauder's coherent-state path
integral for spin~\cite{Klauder79,Stone89},
and the Feynman-Vernon action (after integrating out the bath)
are therefore interference terms between the forward and backward coherent-state paths.
In Sections~\ref{sec:many-spins-many-baths} 
and~\ref{sec:many-spins-one-bath}
I discuss two simple models  
where each qubit is either connected to 
its own environment, or where all the qubits 
are connected to one and the same environment.
In Section~\ref{sec:toric-code} 
I treat Kitaev's toric code interacting with the same kind of environment, and in Section~\ref{sec:discussion} I summarize the paper.
Appendices contain standard material on Klauder's path integral
and the Feynman-Vernon theory, 
an annotated discussion of the pessimistic hypothesis
as formulated in~\cite{Kalai-2016},
and a discussion of quantum error correction and recovery, included for completeness.

\section{Preliminaries and relation to the literature} 
\label{sec:relation-to-literature}
The operation of an ideal quantum computer without decoherence
can be cast in a language similar to information theory 
where the elementary operations (quantum gates) are implemented
as unitary transformations on a pure quantum state, acting on a 
few qubits at a time~\cite{Deutsch89,Barenco95}.  
The errors made in a quantum computation due to decoherence were 
first 
discussed quantitatively by Aharonov, Kitaev and Nisan in~\cite{AharonovKitaevNisan98}
using a formalism where the state is a density matrix and
the transformations are quantum operations mapping density matrices to density matrices.  
This standard formalism is outlined in Section~\ref{sec:pessimistic-hypothesis} below.
In~\cite{AharonovKitaevNisan98} the quantum operations were patterned after 
unitary quantum gates and assumed to factorize both over
operations and between qubits which are not acted upon
by the same operation. The computational power of such a quantum computer
can hence be expressed as properties of networks of ``noisy quantum gates''. 
The main conclusion of~\cite{AharonovKitaevNisan98} 
is that if a quantum computer of this type makes $L$ quantum operations (operates $L$ quantum gates in 
total, counted with multiplicity)
each with error $\epsilon$, then the
combined error of the total operation scales as $L\epsilon$.
This estimate leads to higher accuracy needed the
larger the quantum computer. It therefore points to the need for
quantum error correction to make quantum computation possible~\cite{Kitaev97,Fowler12,Mirrahimi14}.

Quantum or classical error correction must be formulated relative to some
error model. It is for instance trivially impossible to correct 
the errors caused by a transmission channel which completely forgets the initial state.
An important class of error models in quantum information theory, which will be discussed again below in
Sections~\ref{sec:pessimistic-hypothesis} and~\ref{sec:quantum-noise}, assumes that the 
quantum operation has a block structure
where each block acts on the states of one qubit~\cite{DennisKitaevLandahlPreskill2002,Terhal15}.
Such an error model is hence local both 
in space (physical qubits) and in time (no memory).
Furthermore, in the same context it is often assumed that errors are Pauli channels
which describe mis-application of operators.
The analysis of such error models have been generalized 
to probability distribution over channel histories~\cite{TerhalBurkard2005,Bombin2016},
which are therefore no longer local in time, but still local in space.

The originally proposed purpose of quantum computing 
was to simulate another quantum system of interest.
Feynman argued that in general this cannot be done with a classical
device~\cite{Feynman82}; experimental and theoretical
progress in this direction of research was recently reviewed in~\cite{GeorgescuAshhabNori2014}.
Quantum supremacy is a term 
for similar efforts formulated in the language of
computational complexity theory~\cite{Preskill2012,LundBremnerRalph2017,HarrowMontanaro2017}.
The objective is then to solve some classically impossible computational problem
using quantum resources, or to show mathematically that an
output of some quantum device needs many more resources to be simulated classically.
A central model problem has here been BosonSampling~\cite{AaronsonArkhipov2013},
related to computing the permanent of a matrix.
It was shown already in~\cite{TroyanskyTishby1996}
that the permanent appears in an exact expression for the probability of scattering 
bosons, and that it therefore can be estimated by an ideal quantum device.
In contrast, while the determinant of matrix can be computed in polynomial
time \textit{e.g.} by diagonalization, all general classical algorithms to compute a permanent
take exponential time in the size of the matrix.

Noisy BosonSampling is the task of sampling from 
the distribution of a number of non-ideal scattered bosons (photons),
whether or not that distribution is related to a permanent, or 
has some other application~\cite{NikolopoulosBrougham2016}.
Aaronson and Arkhipov showed that Noisy BosonSampling 
remains computationally
hard for somewhat abstract and small noise~\cite{AaronsonArkhipov2013},
while Kalai and Kindler showed that it becomes simple when the scattering
matrix is perturbed by another small but fixed matrix~\cite{KalaiKindler2014}.
Closer to the physics of the problem Aaronson and Brod
showed that Noisy BosonSampling is hard when at most a finite number of
the photons are dropped~\cite{AaronsonBrod2016}
while Oszmaniec and Brod showed that it is easy if out of $n$ photons
all but $\sqrt{n}$ are dropped~\cite{OszmaniecBrod2018}.
It is currently unknown whether Noisy BosonSampling is hard or easy
when a constant fraction ($\alpha n$) of $n$ photons are dropped.

Noisy BosonSampling mainly lies outside of the issues studied in this
paper because photon drop, as a quantum problem, is on the
level of second quantization. The focus in the following will be 
on systems composed of a fixed number of spin-$\frac{1}{2}$ fermions, or ``qubits'', 
as has been the case in most of the quantum information theory literature. 
As briefly reviewed in Section~\ref{sec:quantum-noise}
such systems can also behave noisily in various ways, but for the most
part the computational entities can be assumed to be long-lived,
and the analysis can therefore be carried out on the level of first 
quantization.

\section{Statement of the problem} 
\label{sec:pessimistic-hypothesis}
We assume that there are physical systems with Hilbert space of dimension two
that we call qubits. A pure quantum state on $n$ qubits is a complex ray in a $N=2^n$-dimensional Hilbert space.
A mixed state on the same $n$ qubits is a density matrix $\rho$
which is a non-negative Hermitian operator of unit trace; the set of all density matrices has dimension $N^2-1$.
A quantum operation is a linear map from a set of density matrices
to a set of density matrices which we will write $\Phi$; this is a set of dimension $N^4-N^2$. 
The geometry of density matrices
and general operations of a single qubit ($n=1$, $N=2$) are well understood as 
the Bloch sphere and linear transformations of the Bloch sphere, 
but for higher dimensions there is no such simple picture~\cite{BengtssonZyczkowski}.
 
Let now $n$ qubits start in the pure state $|i>=|i_1,\ldots,i_n>$,
density matrix $\rho_i=|i_1,\ldots,i_n\rangle\langle i_1,\ldots,i_n|$,
and let there be a unitary quantum operation $\Phi_0\rho_i = U\rho_iU^{\dagger}$
with the property that if the qubits are measured in the final state then
the Boolean vector $\mathbf{f}=(f_1,\ldots,f_k)$ is observed with probability 
$P^{(0)}_{if}=\langle f|\Phi_0\rho_i|f\rangle $.
Let then the system be coupled to an environment and described by a quantum operation $\Phi$
and corresponding probabilities $P_{if}=\langle f|\Phi \rho_i|f\rangle$.
A basic measure of the error of
$P_{if}$ with respect to $P^{(0)}_{if}$ is the variational distance between 
the two probability distributions:
\begin{equation}
  \label{eq:TVD-def}
  \hbox{TVD} = \sum_{\mathbf{f}} | P_{if} - P^{(0)}_{if}|
\end{equation}
Any choice of final observable $O$ taking values $o$
leads to probability distributions $P_{io}$ and $P^{(0)}_{io}$,
and it can be shown the maximum of~(\ref{eq:TVD-def})  
over $O$ is the trace norm $\|\cdot  \|_1$ of the difference of the corresponding density 
matrices.
Furthermore, the authors of~\cite{AharonovKitaevNisan98}
introduced the \textit{diamond norm} over 
super-operators and prove the important 
inequalities 
$\| \Phi_1 \Phi_2\|_{\diamondsuit}\leq \| \Phi_1 \|_{\diamondsuit}  \| \Phi_2 \|_{\diamondsuit}$
(Lemma~12, statement 3)
and 
$\| \Phi \rho\|_1 \leq \| \Phi \|_{\diamondsuit}  \| \rho \|_1$
(Lemma~12, combining statements 1 and 2).
For two series of quantum operations that can be written
\begin{equation}
  \label{eq:chained-Phis}
  \Phi=\Phi_L\Phi_{L-1}\cdots\Phi_1\qquad\Phi^{(0)}=\Phi_L^{(0)}\Phi_{L-1}^{(0)}\cdots\Phi_1^{(0)}
\end{equation}
and where each pair of unitary and noisy quantum operations satisfies 
$\| \Phi_l-\Phi_l^{(0)}\|_{\diamondsuit}<\epsilon$ this leads to (\cite{AharonovKitaevNisan98}, Theorem~4)
\begin{equation}
  \label{eq:AKN-Theorem-4}
\hbox{TVD}\leq L\cdot\epsilon
\end{equation}
Equation (\ref{eq:AKN-Theorem-4}) 
says that the total error of a quantum computer scales linearly with the number of
operations $L$ which in many realistic settings would be proportional 
to the number of computational units ($n$) and the duration of the process ($t$).
If an error rate is defined as $\hbox{TVD}$ per $n$ and $t$ then 
(\ref{eq:AKN-Theorem-4}) 
has the interpretation that the error rate per quantum operation 
of a quantum computer is bounded by a constant. In particular it
does not increase with
the number of qubits the quantum computer is operating on. 
Quantum error correction systems are built on
physical (small-scale) qubits which are used to build logical
(larger-scale) qubits on which the quantum computation is done. Under assumptions
that will be critically discussed in Section~\ref{sec:quantum-noise}, the linear scaling
(3) together with sufficiently small error-rate for individual physical qubits and gates acting on them
allows quantum computing based on quantum error-correction.

Kalai's ``pessimistic hypothesis''~\cite{Kalai-2016} 
is the contrary position to the above.
For concreteness I will formulate it as follows:
\begin{quote}
  \textbf{Pessimistic hypothesis:} for a large enough quantum computational system it is not possible to 
  maintain the scaling of the error in (\ref{eq:AKN-Theorem-4}) due to correlations of errors in space and time. 
\end{quote}
If true, this would render high-quality quantum error-correction in particular, and quantum
computing in general impossible for large enough systems.
At this point it must be emphasized that the above statement cannot be found in~\cite{Kalai-2016};
it is a reformulation of the pessimistic hypothesis such that the methods used in this paper are applicable.
I argue in Appendices~\ref{app:Kalai-epsilon} and~\ref{app:scaling-Kalai-epsilon}
that it is a reasonable reformulation,

The problem addressed in this paper is to discuss bounds on the left-hand side of (\ref{eq:AKN-Theorem-4}) 
without considering error rates of individual components at all.
The objective is hence to circumvent the critique of~\cite{Kalai-2016}
by treating the problem as one of the physics of open quantum system,
and not as one of quantum information theory.
The tool to do this, used in Sections~\ref{sec:open-quantum-model}--\ref{sec:toric-code}, 
is the Feynman-Vernon formalism.
As measurements would usually be performed in some
pre-determined way which would often more or less amount to measuring
the $z$-components of all the qubits I will for simplicity 
assume a given initial state $\rho_i$ and a given final
observable, and consider all the variability of the problem 
to stem from $\Phi$ being different from $\Phi^{(0)}$.
The trace norm and the diamond norm will therefore not appear in the following analysis.
Furthermore, error rate is not a concept intrinsic to quantum mechanics.
In the following error rate will therefore only be discussed as an auxiliary 
quantity defined in terms of  $\hbox{TVD}$, in the same way as done above, in text below equation
(\ref{eq:AKN-Theorem-4}).

\section{Quantum noise}
\label{sec:quantum-noise}
As quantum noise is central to the problem addressed in this paper
I will in this Section make a detour and outline the
theoretical and experimental boundaries within which I discuss this
concept. The reader primarily interested
in the main argument may proceed directly to Section~\ref{sec:open-quantum-model}.

Quantum mechanics is based on 
unitary evolution of a state between measurements
and non-unitary collapse of the wave function when it is measured.
The latter is a source of uncertainty which is taken
to be a basic property of the world~\cite{WheelerZurek}.
Quantum computing without decoherence
fully incorporates
this quantum mechanical measurement uncertainty which therefore does not need to be considered further here.

Unitary time evolution shares with Hamiltonian dynamics in classical mechanics 
the property that it is deterministic 
and time-reversal invariant. It is therefore, in 
a colloquial sense of the word, noise-free, and does not, by itself, 
explain  the subjective human experience that time flows 
forward towards the future, and not towards the past.
Modifications of the equations of quantum mechanics
to be stochastic were considered in~\cite{GhirardiRiminiWeber1986}
and more recently discussed by Weinberg~\cite{Weinberg2012,WeinbergNYROB2017}.
The success of quantum mechanics as physical theory implies
that such modifications, if they exist, must be very small.
Such hypothetical modifications can therefore also be ignored in the present context;
Weinberg in~\cite{Weinberg2016} gives a relative bound of $10^{-17}$
by comparing to the stability of atomic clocks.

The issue of quantum noise is instead that
a quantum mechanical system may effectively develop 
in a different manner than by unitary time evolution because
it is interacting with another (unobserved) system.
There are two ways in which this can be described: 
by quantum operations acting on the density matrix
of the system, as summarized above in Section~\ref{sec:pessimistic-hypothesis},
or by explicitly modeling the time evolution of
the observed and unobserved systems together.
It is well known that every quantum operation
has an environmental representation
but that this is not unique; many environments and couplings to the environment
correspond to the same quantum operation on the system~\cite{BengtssonZyczkowski}.   

From the point of view of information theory
the simplest and most natural quantum noise models are the 
quantum operations that are structurally simple and most similar
to unitary evolution.
The factorized error model in~\cite{AharonovKitaevNisan98}
outlined in Section~\ref{sec:pessimistic-hypothesis}
assumes that the elementary unitary transformation of
a noise-free quantum gate is modified to a quantum operation
that acts non-trivially only on these same qubits.
A more physical interpretation was given in~\cite{TerhalBurkard2005}
where each qubit is attached to its own separate environment (a ``bath''),
and then extended to the case where
these baths interact when and in the same combinations as the qubits do.
Other contributions have extended the model and methods of~\cite{AharonovKitaevNisan98}
to when the quantum operations depend on time~\cite{TerhalBurkard2005,P,Bombin2015,Bombin2016}.
All these contributions (and others) have in common a high level of mathematical
sophistication, and the need for assumptions that are physically questionable, or at least not simple.

From the point of view of physics
the simplest and most natural quantum noise models are 
instead those that result from simple interactions
with simple environments. The simplest of these are linear interactions with an environment
of harmonic oscillators. This is the model
that will be introduced in Section~\ref{sec:open-quantum-model} below and used as the basis of the subsequent analysis.
Such models describe a system interacting with delocalized degrees of freedom such as 
photons (in cavity electrodynamics) or phonons (in solid state systems).

It is worth emphasizing that the two views on simplicity are not aligned;
in fact they are more nearly orthogonal.
The quantum operation that results from a harmonic oscillator bath 
originally in a thermal state depends strongly on bath temperature.
If sufficiently high
then the resulting time development of the system is Markovian
\textit{i.e.} factorizes over time, one of the assumptions made in~\cite{AharonovKitaevNisan98}.
In this same limit the system however behaves nearly classically~\cite{BreuerPetruccione,CaldeiraLeggett83a},
not a desired property of a quantum computing device.
If on the other hand bath temperature is low
then the noise from the bath acting on the system will be moderate in overall size 
but correlated in time and space. 
In the regime where a system could work as a quantum computer
it must thus be able to deal with such non-trivial noise,
at least as long as it may be interacting with phonons or photons.

The current leading technology 
for future quantum computers are coupled
superconducting quantum circuits~\cite{DevoretWallrafMartinis,WendinShumeiko}.
Each logical element (qubit) is then 
in fact formed by a mesoscopic object 
containing many millions of atoms, but 
where the behavior of one degree 
of freedom can be assimilated to
that of one quantum spin.
A figure-of-merit of how accurate
is such a description is the ratio
between the gate time of operation
and the qubit relaxation time for
which the current experimental (published) record for coherent super-positions 
is about $5\cdot 10^{-4}$.
This is based on $T_2\sim 20\mu s$ and a previously established cycle time about $10 ns$,
alternatively one can give the number $2\cdot 10^{-4}$
based on the qubit relaxation time $T_1\sim 60\mu s$~\cite{PhysRevLett.107.240501}.
The current (published) record for a system of nine qubits, 
and with all properties measured in the same system, 
is for one qubit (one out of nine) about $10^{-3}$.
This number is based on measured relaxation times $T_1 = 18-41 \mu s$ and 
measured operation times $20-45 ns$, as given in~\cite{Kelly15}, Table S3.

The qubit degree of freedom in the quantum circuit interacts with the other degrees of freedom in 
the circuit, with degrees of freedom in the surrounding device and material, and with as 
external control potential, an influence also mediated by the degrees of freedom of the device.
The total dynamics is hence potentially quite complex.
Deviations from desired dynamics include 
changes in density matrix of computational states of the qubit
as well as leakage, \textit{i.e.}
excitations of higher non-computational states of the qubit.
Considering only the first type of effects
they can be modeled by interactions 
between a qubit and an environment,
the kind of model to be introduced in Section~\ref{sec:open-quantum-model} below.
As recently reviewed 
in~\cite{PaladinoGalperinFalciAltshuler14}, fast environmental modes have to be treated 
quantum mechanically while slow environmental modes can be treated as classical random fields.
The analysis in Section~\ref{sec:open-quantum-model} and following hence pertain
to the fast environmental modes, treated as a harmonic oscillator bath interacting
linearly with the qubit.

The influence of classical random fields on the density matrix of a qubit
will be a a superposition of random unitary transformations
\textit{i.e.} $\rho\to V_i \rho V_i^{\dagger}$, each 
unitary $V_i$ applied with probabilities $p_i$.
On a single qubit all such transformations can be represented as Pauli channels
\textit{i.e.} $\rho\to p_0\rho +\sum_{i}p_i \hat{\sigma}^i \rho \hat{\sigma}^i$ where
$\hat{\sigma}^i$ are the Pauli matrices, and $(p_0,p_1,p_2,p_3)$ are non-negative numbers that sum to one
(take $V_i = e^{i\frac{\pi}{2}\hat{\sigma}_i}=i\hat{\sigma}_i$).
A qubit system perturbed by a Pauli channel is one of the standard models in the quantum computing 
literature~\cite{Kitaev97,DennisKitaevLandahlPreskill2002,P,Fowler12}, and the factorized error model in~\cite{AharonovKitaevNisan98} is obviously also of the same general kind.
As follows from the preceding discussion such models are not realistic
descriptions of interactions with an environment:
as they have no memory the corresponding environmental modes should be treated quantum mechanically.
On the other hand, Pauli channels and similar models describe
the effects on the quantum system of a memory-less classical uncertainty in the control potential. 
Note in passing that the number of independent unitary transformations in $N$-dimensional Hilbert space is $N^2-1$, and
the dimensionality of the class of random super-positions is thus only a $1/N^2$-small fraction of all quantum operations.
For instance, all random superposition of unitary transformations are unital (preserve the identity) 
and therefore do not include \textit{e.g.} amplitude decay channels~\cite{BengtssonZyczkowski}.

The kind of error models considered in the more recent quantum information 
literature which include memory~\cite{TerhalBurkard2005,Bombin2015,Bombin2016} are more 
aligned with the influence of slow environmental modes.
However, $1/f$-noise is an ubiquitous property of solid state devices, and 
this may lead stronger memory effects than have been analyzed up to now,
for further discussion the interested reader is referred to~\cite{PaladinoGalperinFalciAltshuler14}.

Summarizing this Section, quantum noise in systems currently considered for quantum
computing can be classified as (A) classical noise acting quantum mechanically,
(B) influence from a slow quantum environment that can be described classically,
and (C) influence from a fast quantum environment that has to be described quantum mechanically.
From a fundamental point of view only (C) can be an obstacle to quantum, as opposed
to classical, computing. From a practical 
and experimental point of view any of (A), (B) and (C)
could be the main problem. Most of the quantum information literature tacitly assume
(A), and as will be shown in the following analysis it is correct that (C)
generally gives weaker effects than (A).
The effects of (B) are more difficult to treat, likely more system dependent,
and could well be main obstacles to successful quantum computing, as argued  
in~\cite{PaladinoGalperinFalciAltshuler14}. 

\section{The open quantum system model}
\label{sec:open-quantum-model}
The aim of this section is to compare the two probabilities $P_{if}$ and $P^{(0)}_{if}$
in one term in (\ref{eq:TVD-def}) when the quantum computer interacts 
with a heat bath. The Hamiltonian describing the quantum computer
and the bath together is 
\begin{equation}
\label{eq:hamiltonian-spin-bath-model}
\hat{H} = \hat{H}_S + \hat{H}_I + \hat{H}_B
\end{equation}
where $\hat{H}_S$ depends only on the variables describing the quantum computer,
from hereon also referred to as the system,
$\hat{H}_B$ depends only on the bath variables, and $\hat{H}_I$ describes the
interaction of the system and the bath. 
We first consider the system without the heat bath and use
the observation that any unitary transformation can be implemented 
by unitary transformation acting on at most two qubits at a time~\cite{Deutsch89,Barenco95}.  
The system Hamiltonian will thus be
\begin{equation}
\label{eq:pairwise-H}
\hat{H}_S = \hbar \sum_a \mathbf{\mu}_a \mathbf{S}^a + \hbar \sum_{ab} \mathbf{S}^a  \mathbf{\kappa}_{ab} \mathbf{S}^b  
\end{equation}
where $\mathbf{S}^a=\{\hat{S}^a_x,\hat{S}_a^y,\hat{S}_a^z\}$ are the spin operators acting on the $a$'th qubit,
$\mathbf{\mu}_a$ is a 3-vector and $\mathbf{\kappa}_{ab}$ is a 3-by-3 matrix.
Both the $\mathbf{\mu}$'s and the $\mathbf{\kappa}$'s have dimension frequency and can depend on time as required to
implement the overall unitary transformation
\begin{equation}
\label{eq:U-S}
U = {\cal T}e^{-\frac{i}{\hbar}\int \hat{H}_S dt}
\end{equation}
where ${\cal T}$ means time ordering. 
Following the prescription of~\cite{Klauder79} we insert an over-complete
resolution of the identity and write
\begin{widetext}
\begin{eqnarray}
\langle\mathbf{f}|U|\mathbf{i}\rangle&=&\int \prod_a \frac{\sin\theta_a^{(i)} d\theta_a^{(i)}d\phi_a^{(i)}}{2\pi} \frac{\sin\theta_a^{(f)}d\theta_a^{(f)}d\phi_a^{(f)}}{2\pi} \nonumber \\
&& \langle\mathbf{f}|\mathbf{\theta}^{(f)},\mathbf{\phi}^{(f)}\rangle K_{cs}(\mathbf{\theta}^{(f)},\mathbf{\phi}^{(f)},\mathbf{\theta}^{(i)},\mathbf{\phi}^{(i)})   \langle\mathbf{\theta}^{(i)},\mathbf{\phi}^{(i)}|\mathbf{i}\rangle
\label{eq:coherent-state-rep}
\end{eqnarray}
\end{widetext}
where $(\theta^{(i)}_a,\phi^{(i)}_a)$ and $(\theta^{(f)}_a,\phi^{(f)}_a)$ 
parametrize unit spheres,
$K_{cs}$ is the coherent-state propagator and $|\mathbf{\theta}^{(i)},\mathbf{\phi}^{(i)}\rangle$ 
and $|\mathbf{\theta}^{(f)},\mathbf{\phi}^{(f)}\rangle$ are the 
initial and final product coherent states. 
The unitary quantum operation is given by
\begin{equation}
\Phi_0 \rho = U\rho U^{\dagger}
\end{equation}
and the first matrix element we are looking for is 
\begin{widetext}
\begin{eqnarray}
P^{(0)}_{if} &=& \int \prod_a \frac{\sin\theta_a^{(f,F)}d\theta_a^{(f,F)} d\phi_a^{(f,F)}}{2\pi}\frac{\sin\theta_a^{(f,B)} d\theta_a^{(f,B)}d\phi_a^{(f,B)}}{2\pi}
\langle\mathbf{f}|\mathbf{\theta}^{(f,F)},\mathbf{\phi}^{(f,F)}\rangle \langle\mathbf{\theta}^{(f,B)},\mathbf{\phi}^{(f,B)}|\mathbf{f}\rangle \nonumber \\
&& \int \prod_a \frac{\sin\theta_a^{(i,F)}d\theta_a^{(i,F)} d\phi_a^{(i,F)}}{2\pi}\frac{\sin\theta_a^{(i,B)} d\theta_a^{(i,B)}d\phi_a^{(i,B)}}{2\pi}
\langle\mathbf{i}|\mathbf{\theta}^{(i,B)},\mathbf{\phi}^{(i,B)}\rangle \langle\mathbf{\theta}^{(i,F)},\mathbf{\phi}^{(i,F)}|\mathbf{i}\rangle \nonumber \\
&& 
K_{cs}(\mathbf{\theta}^{(f,F)},\mathbf{\phi}^{(f,F)},\mathbf{\theta}^{(i,F)},\mathbf{\phi}^{(i,F)})    
K^*_{cs}(\mathbf{\theta}^{(f,B)},\mathbf{\phi}^{(f,B)},\mathbf{\theta}^{(i,B)},\mathbf{\phi}^{(i,B)})    
\label{eq:forward-backward}
\end{eqnarray}
\end{widetext}
where $F$ means ``forward'' and $B$ means ``backward''.
The coherent-state propagator has a path integral representation 
\begin{equation}
K_{cs}=\lim_{\epsilon\to 0}{\cal N}_{\epsilon} \int \prod_a {\cal D}\theta_a {\cal D}\phi_a e^{i \int 
{\cal L}_{kin} + \epsilon{\cal L}_{reg} + {\cal L}_{S} dt}    
\label{eq:Klauder-path-integral}
\end{equation}
where ${\cal L}_{kin}$ and ${\cal L}_{reg}$ are Klauder's kinetic and regularization terms,
${\cal N}_{\epsilon}$ is a normalization, and ${\cal L}_{S}$ represents the interactions.
The two coherent state propagators in (\ref{eq:forward-backward}) can therefore be written as a double path integral
over forward and backward paths.
I summarize for convenience the Klauder theory in App.~\ref{app:formulae}
where I also give the explicit expression for  ${\cal L}_{S}$ corresponding to
(\ref{eq:pairwise-H}).
For compactness I shall write (\ref{eq:forward-backward}) as
\begin{equation}
P^{(0)}_{if} = \left<\mathbf{1}\right>_{if}
\label{eq:FeynmanVernon}
\end{equation}
where $\left<\cdots\right>_{if}$ is a shorthand the averages implied by  (\ref{eq:forward-backward}) and (\ref{eq:Klauder-path-integral}).

To the above model we now add a bath described by $\hat{H}_B$ and $\hat{H}_I$. The bath 
and the system are originally assumed to be in a product state $\rho^{TOT}_i=\rho_i\oplus \rho_i^{B}$,
and the final total state is $\rho^{TOT}_f=U^{TOT}\rho^{TOT}_i(U^{TOT})^{\dagger}$
where $U^{TOT}$, analogous to $U$ in (\ref{eq:U-S}), depends on the whole Hamiltonian.
The reduced density matrix of the system only at the final time is
$\rho_f=\hbox{Tr}_B[\rho^{TOT}_f]$. The total unitary operator 
$U^{TOT}$ can be represented as a path integral over the forward paths of both the system and the bath,
and analogously for $(U^{TOT})^{\dagger}$.
The initial density matrix of the bath $\rho_i^{B}$ can also be represented as a function of
the starting points of the forward and backward paths of the bath (both to be integrated over). 

As was first shown 
in~\cite{FeynmanVernon} the bath variables can then be integrated 
out. 
Instead of the two coherent-state propagators in (\ref{eq:forward-backward}), each 
expressed as a separate path integral (\ref{eq:Klauder-path-integral}), 
we then instead have
\begin{widetext}
\begin{eqnarray}
K_{FV} &=& \lim_{\epsilon\to 0}|{\cal N}_{\epsilon}|^2 \int \prod_a {\cal D}\mathbf{\theta}^F {\cal D}\mathbf{\phi}^F 
e^{i\int {\cal L}_{kin}(\mathbf{\theta}^F,\mathbf{\phi}^F) + \epsilon{\cal L}_{reg}(\mathbf{\theta}^F,\mathbf{\phi}^F) + {\cal L}_{S}(\mathbf{\theta}^F,\mathbf{\phi}^F)dt} \nonumber \\   
&& \int \prod_a {\cal D}\mathbf{\theta}^B {\cal D}\mathbf{\phi}^B 
e^{-i\int {\cal L}_{kin}(\mathbf{\theta}^B,\mathbf{\phi}^B) - \epsilon{\cal L}_{reg}(\mathbf{\theta}^B,\mathbf{\phi}^B) - {\cal L}_{S}(\mathbf{\theta}^B,\mathbf{\phi}^B)dt} 
e^{i\Phi_{FV}} 
\label{eq:Feynman-Vernon-double-path-integral}
\end{eqnarray}
\end{widetext}
where 
\begin{equation}
\Phi_{FV}=\Phi_{FV}[\mathbf{\theta}^F,\mathbf{\phi}^F,\mathbf{\theta}^B,\mathbf{\phi}^B]
\label{eq:Feynman-Vernon-action}
\end{equation}
is the Feynman-Vernon influence action.
We can then write the second matrix element we are looking for as 
\begin{equation}
P_{if} = \left<e^{i\Phi_{FV}} \right>_{if}
\label{eq:FeynmanVernon-2}
\end{equation}
When the Feynman-Vernon influence action is relatively small we therefore have
\begin{equation}
P_{if}-P_{if}^{(0)} \approx \left<i\Phi_{FV} \right>_{if}
\end{equation}
Extracting a representative value $\overline{\Phi_{FV}}$ we have
\begin{equation}
  \label{eq:TVD-approx}
  \hbox{TVD} = \sum_{\mathbf{f}} | P_{if} - P^{(0)}_{if}|\approx \overline{\Phi_{FV}}
\end{equation}
Eq.~(\ref{eq:TVD-approx}) is the first result of this paper. It means
that the error made by the whole system is determined by a global description
of the system and the environment and is 
proportional to the strength of the interaction between the two.
It therefore allows to estimate the scaling of the error with system size
by estimating the scaling of $\Phi_{FV}$.
In the following two section we will look at two simple models
where this leads to the same scaling as Aharonov-Kitaev-Nisan (eq.~\ref{eq:AKN-Theorem-4}).

\section{The spin-boson model with one bath per spin}
\label{sec:many-spins-many-baths}
The first model of one spin interacting with the environment was the spin-boson model, extensively investigated in~\cite{Leggett87}.
We are here concerned with general interacting spin systems and therefore use a different representation of spin histories than in~\cite{Leggett87}, 
but the description of the bath and the coupling of the system and the bath will be the same.
The model discussed here will hence be referred to as the spin-boson model with one bath per spin (``$1-1$'').
For one spin the terms $\hat{H}_I$ and $\hat{H}_B$ 
in (\ref{eq:hamiltonian-spin-bath-model})
are~\cite{Leggett87}
\begin{widetext}
\begin{equation}
  \label{eq:one-spin-one-bath}
  H_{\hbox{spin-boson}}^{1-1} = \sum_n \hbar\omega_n (a^{\dagger}_n a_n +\frac{1}{2}) + \hat{S}_z \left(\sum_n \sqrt{\frac{\hbar}{2m_n\omega_n}}C_n a_n + \hbox{c.c.}\right) 
\end{equation}
\end{widetext}
where $\hat{S}_z$ is the $z$-component of the spin
and $a^{\dagger}_n$ and $a_n$ and the creation and annihilation operators of harmonic oscillator
labeled by index $n$. The mass and the frequency of the harmonic oscillators are given by $m_n$ and $\omega_n$, and 
the strength of the interaction between the spin and the environment is given by $C_n$. 
For many spins the model discussed in this section assumes one set of terms as in
  (\ref{eq:one-spin-one-bath}) per spin, each with a different set of operators 
and $a^{\dagger}_n$ and $a_n$ 

In the path integral formulation we write  
instead of (\ref{eq:one-spin-one-bath}) the classical Hamiltonian representing the terms involving the environment as
\begin{equation}
    \label{eq:bath-classical}
  H_{\hbox{bath}} = \sum_n \frac{1}{2m_n} p_n^2 + \frac{1}{2}m_n\omega_n^2 x_n^2 + S_z \sum_n C_n x_n  
\end{equation}
where in the coherent-state path integral $S_z$ is the function $\frac{1}{2}\cos\theta$, as discussed in App.~\ref{app:formulae}.
The Feynman-Vernon functional in (\ref{eq:Feynman-Vernon-action}) can then be computed explicitly as a 
functional of the forward and backward spin histories, as outlined in App.~\ref{app:Feynman-Vernon}.
We will here only need the estimate of the Feynman-Vernon action for one spin coupled to one bath given
in (\ref{eq:S_i-S_r-estimate}) and that the Feynman-Vernon actions from more than one disconnected systems
add. The total Feynman-Vernon action in (\ref{eq:Feynman-Vernon-action}) is then estimated as
\begin{equation}
\Phi_{FV}\sim \eta \cdot n\cdot (t_f-t_i) 
\end{equation}
where $n$ is the number of spins, $\eta$ is an overall measure of the strength of the interaction 
between a spin and its bath, and $t_f-t_i$ is the duration of the process.
Following  (\ref{eq:TVD-approx}) and assuming weak coupling (small $\eta$) we then have 
\begin{equation}
  \label{eq:TVD-approx-2}
  \hbox{TVD} \sim \eta \cdot n\cdot (t_f-t_i) 
\end{equation}
Eq.~(\ref{eq:TVD-approx}) is the second result of this paper. It should be read as 
a generalization (\ref{eq:AKN-Theorem-4}) to a definite physical model 
where the interaction strength $\eta$ is what gives rise to the elementary error $\epsilon$, and where 
the number of qubits times the duration of the process ($n\cdot(t_f-t_i) $) plays the
role of of the number of ``noisy operations'' $L$. 
While there are similarities there are also differences. In 
the model used in~\cite{AharonovKitaevNisan98} time does not enter since the system is supposed 
to develop unitarily between the ``noisy operations''. That is a somewhat unphysical assumption as 
any quantum system will interact with the environment to some extent, and therefore 
decohere continuously. 
On the other hand, in the model considered here the complexity 
of the quantum operation that implements the computational task does not enter; all
else equal it does not matter how many operations are performed in the same time
window as long as the form and the strength of the interaction between each qubit and its bath remains the same.

\section{The spin-boson model with one common bath for all spins}
\label{sec:many-spins-one-bath}
A model where each spin has its own bath supposes that each spin is located in a separate material with separate delocalized degrees of freedom.
Although not inconceivable one may also consider the situation where all the spins are located in the same material
and interacting with the same delocalized degrees of freedom.
The interaction terms between the spins and one oscillator in the common bath are then described by the action
\begin{equation}
  \label{eq:bath-common}
  S[x_b,\{\theta_k^f\}] =  \frac{1}{2} \sum_k \int^{t_f}_{t_i} C_{b} x_b(t) \cos \theta_k^f dt   
\end{equation}
where the interaction coefficients $C_{b}$ for simplicity have been taken the same for all spins interacting with the same bath oscillator.
We can re-write the right-hand side of (\ref{eq:bath-common}) as $\frac{n}{2} \int^{t_f}_{t_i} dt C_{n} x_n(t)  \overline{\cos\theta^f}(t)$
where $n$ is the number of spins (qubits) and 
$\overline{\cos\theta^f}=\frac{1}{n}\sum_{k} \cos\theta_k^f$, and same for the backward path, and then integrate out the
bath oscillators. The result will be a Feynman-Vernon influence functional of the
two collective coordinates
of the same structure as (\ref{eq:S_i}) and (\ref{eq:S_r}), and which can be written
\begin{equation}
  \label{eq:FV-common}
  \Phi = n^2 \frac{i}{\hbar} S_i[\overline{\cos\theta^f},\overline{\cos\theta^b}]-n^2\frac{1}{\hbar} S_r[\overline{\cos\theta^f},\overline{\cos\theta^b}]
\end{equation}
Formally (\ref{eq:FV-common}) scales quadratically with number of spins (qubits).
However, it is physically reasonable that an 
increasing number of spins in the same material would take more place.
One may think of either the spins are arranged along a line, or arranged on two-dimensional grid.
In the first case one dimension of the system increases proportional to $n$ while in the second case two dimensions of the system
increase proportional to $\sqrt{n}$, and in both cases 
the interaction coefficient $C$ between the isolated spin and a delocalized mode can be expected to scale as $1/n$.
The number of modes in a small frequency interval $d\omega$ will increase as $n$ and the overall bath power spectrum $J(\omega)$ therefore decreases as $1/n$.
Combining these estimates one gets back the linear scaling in (\ref{eq:TVD-approx-2}).
Furthermore, the differences between the forward and backward paths are fluctuating quantities and at least in the
high-temperature near-classical regime investigated in~\cite{CaldeiraLeggett83a} one can expect
$\overline{\cos\theta^f}-\overline{\cos\theta^b}$ to scale as $1/\sqrt{n}$. The real part of $\Phi$ in (\ref{eq:FV-common})
would therefore give a contribution independent of the number of qubits
while the imaginary part of $\Phi$ in (\ref{eq:FV-common}) would give an error increasing slower than linearly.

\section{The toric code in the Feynman-Vernon theory}
\label{sec:toric-code} 
A canonical  model of quantum computing and quantum error correction is Kitaev's toric code~\cite{Kitaev97}.
In the simplest version, which will be considered here, an $N\times M$ lattice of spins 
are located at edges in a regular lattice on the 2-torus, and operated on by operators
called stabilizers
\begin{equation}
\label{eq:Kitaev-operators}
A_s = \prod_{i\in\hbox{star}(i)}\sigma_i^x\qquad B_p = \prod_{i\in \partial p}\sigma_i^z
\end{equation}
In $A$-type stabilizers
$i\in\hbox{star}(i)$ denote the spins (edges) in the neighborhood of a vertex $s$ and
in $B$-type stabilizers $i\in\partial  p$ denote
the spins (edges) around a plaquette $p$; $\sigma_i^z$ and $\sigma_i^x$ are Pauli operators acting
on spin $i$. All the stabilizers commute and the
eigenspace of all of them measured simultaneously is four-dimensional.
This Hilbert space can be identified with that of two spins, usually in this context 
called logical qubits, and Pauli operators on these two qubits
are products of operators on the physical spins taken around the two basic circuits on the torus.
Note that we are here concerned with the Kitaev code, and not the closely 
related quantum statistical mechanical system known as the Kitaev model.
In that second case, see Eq.~\ref{eq:Kitaev-operators} below, the operators 
in Eq.~\ref{eq:Kitaev-operators} are terms in a Hamiltonian operator, and not measured continuously.
The four-dimensional ground state of the Kitaev model is the one where the eigenvalues of 
$A_s$ and $B_p$ in Eq.~\ref{eq:Kitaev-operators} are all equal to one.

More complex versions of toric codes which can accommodate many more qubits will not
be considered further here, nor the very considerable experimental challenges of actually
building such systems; for a recent review, see~\cite{Fowler12}.
The system under consideration hence consists of $NM$ physical spins and a bath of harmonic oscillators
developing according to (\ref{eq:hamiltonian-spin-bath-model}) where in addition the stabilizers
are continuously measured. The system Hamiltonian is thus
\begin{equation}
\label{eq:Kitaev-system-Hamiltonian}
H_S = H_S(s_1^x,s_1^y,s_1^z,s_2^x,s_2^y,s_2^z)
\end{equation}
where 
\begin{eqnarray}
\label{eq:Kitaev-operators}
s_1^x = \prod_{i\in C_1} \sigma_i^x\quad s_1^z = \prod_{i\in C_2'} \sigma_i^z &\quad& s_1^y=is_1^zs_1^x \nonumber \\ 
s_2^x = \prod_{i\in C_2} \sigma_i^x\quad s_2^z = \prod_{i\in C_1'} \sigma_i^z &\quad& s_2^y=is_2^zs_2^x \nonumber 
\end{eqnarray}
are the Pauli operators acting on the logical qubits, and $C_1,C_2$ and $C_1',C_2'$  
are the two basic cycles of the torus in  respectively
the vertex-centered and plaquette-centered lattice. 

A basis of the states of the physical spins is $|i_1,i_2,\ldots,i_{NM}\rangle$ where $i_p=\pm 1$
denotes the up (down) state of spin $p$. An alternative basis is by above given by the 
$k$ values of the $z$-components of logical qubits $l_r$ (here $k=2$)
and the $NM-k$ values 
of the stabilizers $m_q=\pm 1$. 
These two bases are related by a unitary transformation
\begin{equation}
\label{eq:A-transformation}
|\mathbf{l},\mathbf{m}>=\sum_{\mathbf{i}}A^{\mathbf{l},\mathbf{m}}_{\mathbf{i}}|\mathbf{i}>
\end{equation}
where $|\mathbf{l},\mathbf{m}>$ denotes $|l_1,\ldots,l_k,m_1,\ldots,m_{NM-k}>$
and $|\mathbf{i}>$ denotes $|i_1,i_2,\ldots,i_{NM}>$.
By orthogonality of the states of the stabilizers and the logical qubits we have
\begin{equation}
\label{eq:A-orthogonality}
\sum_{\mathbf{i}} A^{\mathbf{l},\mathbf{m}}_{\mathbf{i}} \left(A^{\mathbf{l'},\mathbf{m'}}_{\mathbf{i}}\right)^* 
= \mathbf{1}_{l,l'}\mathbf{1}_{m,m'}
\end{equation}
Now assume that over some stretch of time the measured values of all the stabilizers are constant.
These are then not histories of quantum variables but known classical (and constant) records.
The interaction of the physical spins with one bath oscillator $b$ gives an interaction Hamiltonian for the logical qubits 
 \begin{eqnarray}
\label{eq:Q}
Q(\mathbf{l},\mathbf{l'};\mathbf{m})&=&\langle\mathbf{l'},\mathbf{m}|\sum_r\hat\sigma_r^z|\mathbf{l},\mathbf{m}\rangle \nonumber \\
&=&\sum_{\mathbf{i}}A^{\mathbf{l,m}}_{\mathbf{i}}\left(A^{\mathbf{l',m}}_{\mathbf{i}}\right)^*\left(\sum_r(-1)^{i_r}\right) 
\end{eqnarray}
From these follow interaction Hamiltonians for the coherent-state representations
of the histories of the logical qubits in the forward and backward paths 
 \begin{eqnarray}
\label{eq:Q-FandB}
Q^F(\mathbf{\theta^F},\mathbf{\phi^F};\mathbf{m})&=&\sum_{\mathbf{l},\mathbf{l'}}\langle\mathbf{\theta^F},\mathbf{\phi^F}|\mathbf{l'}\rangle
Q(\mathbf{l},\mathbf{l'};\mathbf{m}) \left<\mathbf{l} | \mathbf{\theta^F},\mathbf{\phi^F}\right> \nonumber \\
Q^B(\mathbf{\theta^B},\mathbf{\phi^B};\mathbf{m})&=&\sum_{\mathbf{l},\mathbf{l'}}\langle\mathbf{\theta^B},\mathbf{\phi^B}|\mathbf{l'}\rangle
Q(\mathbf{l},\mathbf{l'};\mathbf{m}) \langle\mathbf{l} | \mathbf{\theta^B},\mathbf{\phi^B}\rangle \nonumber
\end{eqnarray}
These more complicated functions $Q^F$ and $Q^B$ play the same role for the interaction of the toric
code with a bath of oscillators as the sums of the cosines 
in the simple model discussed in Section~\ref{sec:many-spins-one-bath} above, compare Eq.~(\ref{eq:bath-common}).

The influence functional is as above estimated as
\begin{equation}
\label{eq:FV-Kitaev}
\Phi_{FV}\sim \eta\overline{Q}^2  (t_f-t_i) 
\end{equation}
where $\overline{Q}$ is a typical value of $Q^F$ and $Q^B$. 
A rough estimate of $\overline{Q}$ follows from assuming that 
each element of $A$ in (\ref{eq:A-transformation})
is about $2^{-\frac{NM}{2}}$ with a fluctuating sign,
which is consistent with (\ref{eq:A-orthogonality}). 
$Q(\mathbf{l},\mathbf{l'};\mathbf{m})$ in (\ref{eq:Q}) is then the sum of $2^{NM}$ terms of fluctuating
signs, each of size about $2^{-NM}$ and hence of overall typical size $2^{-\frac{NM}{2}}$. Each of the 
two functions $Q^F$ and $Q^B$ is then a sum of $(2^k)^2$ such terms 
multiplied by the matrix elements with the angles which each have RMS average $2^{-\frac{k}{2}}$ (see Appendix~\ref{app:formulae}).
The approximate sizes of $Q^F$ and $Q^B$ are hence
$2^{\frac{k-NM}{2}}$, the amplitude  $\overline{Q}^2$ in (\ref{eq:FV-Kitaev}) is consequently $2^{k-NM}$, and the influence of
the bath on the states of the logical qubits exponentially small in system size.

A more systematic estimate of $\overline{Q}$ follows from observing that $Q(\mathbf{l},\mathbf{l'};\mathbf{m})$
is the matrix element of the operators coupling the system to the heat bath between two eigenstates  
the Kitaev model given by Hamiltonian
\begin{equation}
\label{eq:Kitaev-operators}
H_K = -\sum_s A_s - \sum_p B_p 
\end{equation}
The two states have the same quantum numbers ($\mathbf{m})$ determined by the 
eigenvalues of the operators $A_s$ and $B_p$, and the same or different quantum numbers
given by the logical operators acting on the logical qubits ($\mathbf{l}$ and $\mathbf{l'}$).
It is known that the matrix elements of local operators
in the ground state of the Kitaev model are exponentially small in system size~\cite{BravyiHastings}.
For $(\mathbf{m})=(1,1,1,\ldots,1)$ the logical qubits of the toric code
are therefore almost insensitive to interactions with bath.

The results of this section are positive for the Kitaev code, and it may be
useful to compare other results in the literature. 
First, the standard view is 
that the Kitaev \textit{model} in 2D with Hamiltonian (\ref{eq:Kitaev-operators}) does
not preserve its state when interacting with a finite-temperature heat 
bath~\cite{DennisKitaevLandahlPreskill2002,AlickiFannesHorodecki2007,BravyiTerhal,AlickiFannesHorodecki,Bombin2013},
a result
often stated as that the Kitaev model is not a stable quantum memory. 
That is not the same setting as considered here,
as the stabilizer operators are then not continuously measured.
To reproduce these results in the formalism of the present paper
one should promote the measured values of the stabilizers $\textbf{m}$
to be quantum variables represented in a larger coherent-state path
integral by forwards and backwards angles $\{\theta^{\textbf{m},F}\phi^{\textbf{m},F}\}$
and $\{\theta^{\textbf{m},B}\phi^{\textbf{m},B}\}$.
Instead of (\ref{eq:Q}) we then have ($\textbf{m}$ and  $\textbf{m'}$ different)
 \begin{eqnarray}
\label{eq:Q-model}
Q(\mathbf{l},\mathbf{m},\mathbf{l'},\mathbf{m'})&=&\langle\mathbf{l'},\mathbf{m'}|\sum_r\hat\sigma_r^z|\mathbf{l},\mathbf{m}\rangle \nonumber \\
&=&\sum_{\mathbf{i}}A^{\mathbf{l,m}}_{\mathbf{i}}\left(A^{\mathbf{l',m'}}_{\mathbf{i}}\right)^*\left(\sum_r(-1)^{i_r}\right) 
\end{eqnarray}
which we can again estimate as $2^{-\frac{NM}{2}}$. The two terms $Q^F$ and $Q^B$ are however now sums of
$(2^{NM})^2$ terms, and are therefore not small in system size. 
In this case estimate (\ref{eq:FV-Kitaev}) hence gives essentially the same result as (\ref{eq:TVD-approx-2}).
%

\section{Discussion}
\label{sec:discussion}
In this work I have considered the error made by a quantum computer weakly coupled to an environment
such that the quantum computer cannot be meaningfully described as a network of ``noisy quantum gates''. 
I have instead estimated the error by
combining Klauder's path integral for spin and a Feynman-Vernon elimination of a thermal bath modeled
as a set of harmonic oscillators interacting linearly with the qubits.

I have looked at three models. In the first two all qubits are computational units and all
interact directly with a heat bath as in the spin-boson model~\cite{Leggett87}.
In these two simpler models no error correction was considered: the goal was to see if the
scaling of the overall error found by   
Aharonov, Kitaev and Nisan in~\cite{AharonovKitaevNisan98} needs to be modified.
The answer is negative. Instead of the error rate of a noisy quantum
gate, a concept not defined for these models, the crucial parameter is the interaction strength
between the system and the heat bath. If that parameter is small the
total error scales at most linearly with system size (number of qubits) and time of operation --
without any assumptions on locality in space and time. 

The third model considered in the toric code of Kitaev~\cite{Kitaev97} in 2D 
where additionally
the physical qubits interact with a heat bath as in the spin-boson model.
The computational units (logical qubits) of this model are non-localized degrees of
freedom, much fewer in number than the physical qubits.
The analysis brings out the fact that the states of the logical qubits are almost insensitive to interactions with a bath, at least
in the ground state of the related Kitaev model where 
all the stabilizers (defined above) have value one.
One consequence of this observation is that such an influence does not need to
be corrected, as it is exponentially small in the system size.
The combination of Klauder's path integral and Feynman-Vernon
allows to treat together the interaction with a heat bath and other errors 
that can be modeled as Pauli channels, and can hence be considered an alternative to the quantum semi-group dynamics 
(Davies generator formalism) within which many systematic studies of this and the related Kitaev model
have be performed previously~\cite{AlickiFannesHorodecki2007,AlickiFannesHorodecki,Bombin2013}.

A thermal bath consisting of harmonic oscillators is a model of delocalized
environmental modes such as phonons. The main degrees of freedom in a real material at very low temperature, such as
defects and nuclear spins, are on the other hand likely to be localized, and may be more 
accurately described as a spin bath~\cite{ProkofevStamp2000}. For this case it may be argued
that the environment of each qubit consists in a finite set of neighboring spins the effects
of which would in principle also be given by a Feynman-Vernon action as in  
(\ref{eq:Feynman-Vernon-action}). Although precise estimates of this action would be more difficult to obtain, there seems to be no 
reason to assume that the number of environmental spins 
interacting with one qubit scales with the number of qubits of the quantum computer. 

Finally, although the analysis is this paper has shown that fast environmental modes
that have to be treated quantum mechanically are not a fundamental
problem for quantum computing, there remains slow environmental modes.
As long as these may be treated classically they cannot be a problem
for quantum computing per se, but may nevertheless still pose very significant obstacles
in practice, a point of view forcefully argued in~\cite{PaladinoGalperinFalciAltshuler14}. 

\section*{Acknowledgments}
I thank Gil Kalai for a stimulating discussion and  Yuri Galperin,
Benjamin Huard, Petteri Kaski, Cris Moore,
Sorin Paraoanu, Sergey Pershoguba and Karol \.{Z}yczkowski
for constructive remarks.
This research has been supported by the Swedish Science Council through grant 621-2012-2982,
by the Academy of Finland through its Center of Excellence COIN, and by the
Chinese Academy of Sciences CAS President's International Fellowship Initiative (PIFI) Grant No. 2016VMA002.

\appendix
\section{A quantum-mechanical formulation of Kalai's $\epsilon$}
\label{app:Kalai-epsilon}
The purpose of this appendix is to argue that
the total variational distance as defined above in (\ref{eq:TVD-def}),
counted per qubit, is a reasonable quantum
mechanical interpretation of the error rate $\epsilon$ discussed in~\cite{Kalai-2016}.
I emphasize that this interpretation can not be found~\cite{Kalai-2016},
but is introduced here as a way to state the problem within the theory of open quantum systems.

To do so we consider the special case where the quantum operation $\Phi_0$,
determined by a unitary transformation $U$, is such that 
there is a single final state $|f\rangle$ with $\langle f|\Phi_0\rho_i |f\rangle =1$.
Applying the quantum operation $\Phi_0$ to $|i\rangle $ can then be said to yield $|f\rangle $ with certainty, and $\Phi_0$ can then be called ``noise-less''.
Applying the ``noisy'' quantum operation $\Phi$ and measuring all the qubits would on the other hand give
the Boolean vector $\mathbf{f}$
with probability $1-\epsilon'$ for some $\epsilon'>0$ and a result different from $\mathbf{f}$ with total probability $\epsilon'$.
Let now further $\Phi$ be such that the probabilities $p(\mathbf{\tilde{f}})$ are sensibly different from zero
only when the Hamming distance between $\mathbf{\tilde{f}}$ and $\mathbf{f}$ is at most one, \textit{i.e.} when at most
one qubit has been flipped, and let the probability to flip any one qubit be $\epsilon=\epsilon'/n$.
The error rate so defined is then the same as $\frac{1}{2}\hbox{TVD}$, where $\hbox{TVD}$ is defined in (\ref{eq:TVD-def}).

\section{Kalai's pessimistic hypothesis}
\label{app:scaling-Kalai-epsilon}
The purpose of this appendix 
is to argue that Kalai's pessimistic hypothesis claims that
$\epsilon$ as introduced above in App.~\ref{app:Kalai-epsilon}
scales linearly with number of qubits in the quantum computer.
The argument proceeds by selected quotes from~\cite{Kalai-2016}.
We start from
\begin{quote}
``The error rate in every realization
of a universal quantum circuit scales up (at least) linearly
with the number of qubits'
\end{quote}
Readers of~\cite{Kalai-2016} will note that this statement is followed by
\begin{quote} 
 ``The effort required to obtain a
bounded error level for any implementation of universal
quantum circuits increases (at least) exponentially with
the number of qubits''
\end{quote}
which is also important
to Kalai's argument concerning universal quantum computers. In the present discussion,
which focuses on the consequences for open quantum systems,
I will however limit myself to the first part.

In~\cite{Kalai-2016} Kalai also argues by the example of a depolarizing
one-qubit channel described by
\begin{equation}
  \label{eq:depolarizing}
  \Phi\rho = (1-p) \rho + p \frac{1}{2}\mathbf{1}
\end{equation}
where $\rho$ is the density matrix of a qubit (a positive Hermitian $2$-by-$2$ matrix of unit trace), $\Phi$ is the quantum operation
(a linear operator of the set of such matrices on itself) and
$\frac{1}{2}\mathbf{1}$ is the completely depolarized density matrix. 
The error rate is then taken to be $p$ and, more generally
\begin{quote}
``...error rate can be defined as the probability that a
qubit is corrupted at a computation step, conditioned on it surviving up to this step''
\end{quote}
which is followed by
\begin{quote}
``...when we say that the rate
of noise per qubit scales up linearly with the number
of qubits, we mean that when we double the number of qubits in the circuit, 
the probability for a single qubit
to be corrupted in a small time interval doubles''
\end{quote}
In combination the above quotes imply that Kalai's pessimistic hypothesis states that the total
error of the whole system scales at least quadratically with the number of qubits. 
In the interpretation used here, see App.~\ref{app:Kalai-epsilon} above, this is
taken to mean that the total variational distance in (\ref{eq:TVD-def}) also scales at least quadratically.
Similarly too the main text and App.~\ref{app:Kalai-epsilon} I emphasize again 
that this quadratic global scaling cannot be found~\cite{Kalai-2016}
but is a consequence of the further interpretations introduced here.

\section{The Klauder coherent-state path integral for spin}
\label{app:formulae}
This appendix summarizes properties pertaining to the Klauder coherent-state path integral. The coherent states are defined as
\begin{equation}
\label{eq:coherent-states-def}
|\theta,\phi\rangle =e^{-i \phi\hat{S}_z} e^{-i\theta\hat{S}_y} |\uparrow \rangle=\left(\begin{array}{c}e^{-\frac{i}{2}\phi}\cos\frac{\theta}{2}\\ e^{\frac{i}{2}\phi}\sin\frac{\theta}{2}\end{array}\right)
\end{equation} 
The two angles $\theta\in [0,\pi]$ and  $\phi\in [0,2\pi]$ parametrize the unit sphere with area $4\pi$.
The various matrix elements used in the main text and below are hence
\begin{equation}
\label{eq:coherent-def-matrix-elements}
\langle\uparrow|\theta,\phi\rangle =e^{-\frac{i}{2}\phi}\cos\frac{\theta}{2}\quad \langle\downarrow|\theta,\phi\rangle=e^{\frac{i}{2}\phi}\sin\frac{\theta}{2}\nonumber
\end{equation}
and the matrix element between two coherent states is
\begin{widetext}
\begin{eqnarray}
\langle\theta',\phi'|\theta,\phi\rangle &=&  \langle\theta',\phi'|\uparrow\rangle \langle\uparrow|\theta,\phi\rangle+ \langle\theta',\phi'|\downarrow\rangle\langle\downarrow|\theta,\phi\rangle \nonumber \\  
                            &=&\cos\frac{\phi'-\phi}{2}\cos\frac{\theta'-\theta}{2} + i\sin\frac{\phi'-\phi}{2}\cos\frac{\theta'+\theta}{2} \nonumber 
\end{eqnarray}
\end{widetext}
When the two sets of angles are close this matrix element is
\begin{eqnarray}
\label{eq:approximate-matrix}
<\theta',\phi'|\theta,\phi> &\approx&  1 +\frac{i}{2}(\phi'-\phi)\cos\theta 
\end{eqnarray}
up to terms which are small as $(\phi'-\phi)^2$ and $(\theta'-\theta)^2$. 
Matrix elements of the operator for the $z$-component of spin are
\begin{eqnarray}
\label{eq:S_z}
\langle\theta',\phi'|\hat{S}_z | \theta,\phi\rangle &=&   \langle\theta',\phi'|\uparrow\rangle\langle\uparrow|\hat{S}_z|\uparrow\rangle \langle\uparrow|\theta,\phi\rangle+ \nonumber \\
                                       &&  \langle\theta',\phi'|\uparrow\rangle\langle\uparrow|\hat{S}_z|\downarrow\rangle \langle\downarrow|\theta,\phi\rangle+ \nonumber \\
                                       &&  \langle\theta',\phi'|\downarrow\rangle\langle\downarrow|\hat{S}_z|\uparrow\rangle \langle\uparrow|\theta,\phi\rangle+ \nonumber \\
                                       &&  \langle\theta',\phi'|\downarrow\rangle\langle\downarrow|\hat{S}_z|\downarrow\rangle \langle\downarrow|\theta,\phi\rangle \nonumber \\
                                       &=&  \frac{1}{2}(\cos\frac{\phi'-\phi}{2}\cos\frac{\theta'+\theta}{2} + \nonumber \\
                                       &&\qquad i\sin\frac{\phi'-\phi}{2}\cos\frac{\theta'-\theta}{2})
\end{eqnarray}
which when the two sets of angles are close means
\begin{eqnarray}
\label{eq:approximate-S_z}
\langle\theta',\phi'|\hat{S}_z | \theta,\phi\rangle &\approx&   \frac{1}{2}\cos\theta 
\end{eqnarray}
Similarly
\begin{eqnarray}
\label{eq:approximate-S}
\langle\theta',\phi'|\hat{S}_x | \theta,\phi\rangle &\approx&   \frac{1}{2}\sin\theta\cos\phi \\
\label{eq:approximate-S-2}
\langle\theta',\phi'|\hat{S}_x | \theta,\phi\rangle &\approx&   \frac{1}{2}\sin\theta\sin\phi
\end{eqnarray}
The vector $\vec S=\frac{1}{2}\left(\sin\theta\cos\phi,\sin\theta\sin\phi,\cos\theta\right)$ is the radial vector of length $\frac{1}{2}$,
polar angle $\theta$ and azimuthal angle $\phi$.
The coherent states provide a partition of the unity in an over-complete basis.
Using (\ref{eq:coherent-def-matrix-elements}) we have 
\begin{eqnarray}
\langle\uparrow|\uparrow\rangle &=& \int \frac{\sin\theta d\phi d\theta}{2\pi} \langle\uparrow|\theta,\phi\rangle  \langle\theta,\phi|\uparrow\rangle = 1 \nonumber \\
\langle\uparrow|\downarrow\rangle &=& \int \frac{\sin\theta d\phi d\theta}{2\pi} \langle\uparrow|\theta,\phi\rangle  \langle\theta,\phi|\downarrow\rangle = 0 \nonumber \\
\langle\downarrow|\uparrow\rangle &=& \int \frac{\sin\theta d\phi d\theta}{2\pi} \langle\downarrow|\theta,\phi\rangle  \langle\theta,\phi|\uparrow\rangle = 0 \nonumber \\
\langle\downarrow|\downarrow\rangle &=& \int \frac{\sin\theta d\phi d\theta}{2\pi} \langle\downarrow|\theta,\phi\rangle \langle\theta,\phi|\downarrow\rangle = 1 \nonumber 
\end{eqnarray} 
A time evolution operator can therefore be expressed as
\begin{eqnarray}
{\cal T}e^{-\frac{i}{\hbar}\int \hat{H} dt} &=& \prod_n \int \frac{\sin\theta_n d\phi_n d\theta_n}{2\pi}\cdots |\theta_{n+1},\phi_{n+1}\rangle\nonumber \\
                                    && \langle\theta_{n+1},\phi_{n+1}|e^{-\frac{i}{\hbar}\int_{t_n}^{t_{n+1}} \hat{H} dt} |\theta_{n},\phi_{n}\rangle\nonumber \\
                                    &&\langle\theta_{n},\phi_{n}|\cdots
\label{eq:time-evolution}
\end{eqnarray} 

If it can be arranged that two consecutive sets of angles are close, the interaction term  (\ref{eq:Klauder-path-integral}) for the 
interaction Hamiltonian (\ref{eq:pairwise-H}) is, using (\ref{eq:approximate-S_z}) and (\ref{eq:approximate-S}),
\begin{eqnarray}
\label{eq:pairwise-L}
{\cal L}_{S} &=& -\frac{1}{2}\sum_a \mu_a^z\cos\theta_a+(\mu_a^x\cos\phi_a+\mu_a^y\sin\phi_a) \nonumber \\
&& - \frac{1}{4}\sum_{ab} \kappa_{ab}^{zz}\cos\theta_a \cos\theta_b + \cos\theta_a \sin\theta_b(\kappa_{ab}^{zx}\cos\phi_b \nonumber \\  
&& +\kappa_{ab}^{zy}\sin\phi_b)  + (\kappa_{ab}^{xz}\cos\phi_a+\kappa_{ab}^{yz}\sin\phi_a)\sin\theta_a\cos\theta_b  + \nonumber \\
&& \quad \sin\theta_a\sin\theta_b(\kappa_{ab}^{xx}\cos\phi_a\cos\phi_b \kappa_{ab}^{xy}\cos\phi_a\sin\phi_b +  \nonumber \\ 
&& \quad \kappa_{ab}^{yx}\sin\phi_a\cos\phi_b  + \kappa_{ab}^{yx}\sin\phi_a\cos\phi_b + +\mu_a^y\sin\phi_a) \nonumber \\
&& \qquad (\mu_b^x\cos\phi_b+\mu_b^y\sin\phi_b)
\end{eqnarray}
where the factor $\hbar$ has been included for convenience, compare 
(\ref{eq:pairwise-H}) and (\ref{eq:Klauder-path-integral}).
For the discussion below and in the main text it only matters that (\ref{eq:pairwise-L}) is some definite function of the angles parametrizing the spin history.

To enforce that two consecutive sets of angles are close one uses
a regularization term 
\begin{equation}
\label{eq:regularization}
{\cal L}_{reg}=\frac{1}{2}\left(\dot{\theta}^2+\sin^2\theta\dot{\phi}^2)\right)
\end{equation}
The matrix element (\ref{eq:approximate-matrix})
can then be written 
\begin{eqnarray}
\label{eq:approximate-matrix-2}
\langle\theta',\phi'|\theta,\phi\rangle &\approx&  e^{i\int_{t}^{t'} {\cal L}_{kin}} 
\end{eqnarray}
defining the kinetic term in (\ref{eq:Klauder-path-integral}):
\begin{equation}
\label{eq:kinetic}
{\cal L}_{kin}=\frac{1}{2}\cos\theta\dot{\phi}
\end{equation}
The regularization and interaction terms in (\ref{eq:Klauder-path-integral}) are as in (\ref{eq:regularization})
and (\ref{eq:pairwise-L}) above.
The canonical momenta conjugate to $\theta$ and $\phi$ are
\begin{equation}
p_{\theta} = \epsilon \dot{\theta}\qquad p = p_{\phi}=\epsilon\sin^2\theta \dot{\phi} + \frac{1}{2}\cos\theta
\end{equation}
As discussed in~\cite{Klauder79} and in~\cite{atland-simons}, when $\epsilon$ is set to zero
the remaining action is first order. The function $\frac{1}{2}\cos\theta$ then takes the meaning of momentum $p$, conjugate to $\phi$.
and the radial vector $\vec S$ can be written 
$\left(\frac{1}{2}\sqrt{1-4p^2}\cos\phi,\frac{1}{2}\sqrt{1-4p^2}\sin\phi,p\right)$.
The Poisson brackets of the components of this vector 
satisfy the angular momentum relations. This suggests that in the path integral
the operators $\hat{S}_x$, $\hat{S}_y$ and $\hat{S}_z$ should be translated into these
functions $S_x$, $S_y$ and $S_z$, which is indeed the prescription given by (\ref{eq:approximate-S_z}), (\ref{eq:approximate-S}),
(\ref{eq:approximate-S-2}) and (\ref{eq:pairwise-L}).

With the regularization (\ref{eq:regularization}) the path integral is a standard one, and $\epsilon$ could be interpreted as the 
mass of a particle confined to move on the surface of a sphere of fixed radius. The path integral in curved space (as is the sphere) is 
a well-developed topic with several complexities~\cite{Schulman}, but for the present discussion, where the ``mass term'' is only
for regularization, one can simply interpret the integral of (\ref{eq:regularization}) as a time discretization
\begin{equation}
\label{eq:regularization-discretization}
\epsilon\int_{t_i}^{t_f}{\cal L}_{reg}\approx \sum_n \frac{\epsilon}{2\Delta t_n}\left(\Delta\theta_n^2+\overline{\sin^2\theta}\Delta\phi_n^2\right)
\end{equation}
and where $\overline{\sin^2\theta}$ indicates \textit{e.g.} mid-point prescription and the normalizing coefficient is
\begin{equation}
\label{eq:N-epsilon}
N_{\epsilon} = \prod_n (\overline{\sin^2\theta_n})^{\frac{1}{2}}\frac{\epsilon}{2\pi i\Delta t_n}
\end{equation}
The regularization and the normalization are a weight on the Fourier components of the spin history, and the path integral built on
a discretization of (\ref{eq:kinetic}) and (\ref{eq:regularization}) can therefore be written 
\begin{equation}
\int \prod_k d\mu_{\epsilon}(\hat{\theta}_k,\hat{\phi}_k) e^{i\int {\cal L}_{kin}[\{\hat{\theta}_k,\hat{\phi}_k\}]}
\end{equation}
For finite $\epsilon$ this weight penalizes high Fourier components. Consider two realizations $\{\hat{\theta}_k,\hat{\phi}_k\}$
and  $\{\hat{\theta'}_k,\hat{\phi'}_k\}$ which coincide at the two endpoints. The difference of their actions 
\begin{equation}
\int {\cal L}_{kin}[\{\hat{\theta'}_k,\hat{\phi'}_k\}] - \int {\cal L}_{kin}[\{\hat{\theta}_k,\hat{\phi}_k\}]
\label{eq:action-diff}
\end{equation}
is the integral of ${\cal L}_{kin}$ around a closed path, which in turn equals the area on the surface of the sphere circumscribed by that path~\cite{atland-simons}.
This area depends only weakly on high Fourier components and the limit of zero $\epsilon$ is therefore well-behaved.

\section{The Feynman-Vernon method for spin histories}
\label{app:Feynman-Vernon}
The starting point is the interaction and bath Hamiltonian for a single spin interacting with a bath:
\begin{equation}
    \label{eq:bath-classical}
  H_{\hbox{bath}}+  H_{\hbox{I}} = \sum_n \frac{1}{2m_n} p_n^2 + \frac{1}{2}m_n\omega_n^2 \left(x_n + S_z \frac{C_n}{m_n\omega_n^2}\right)^2  
\end{equation}
In above $S_z$ is read $\frac{1}{2}\cos\theta$, the function representing the operator $\hat{S}_z$ in the Klauder path integral.
The last term from expanding the squares in (\ref{eq:bath-classical}), $\sum_n S_z^2 \frac{1}{2}\frac{C_n}{m_n\omega_n^2}$, only depends
on the spin history, and is a counter-term which
it has become customary to include in the interaction term~\cite{CaldeiraLeggett83a}.
The two variables $x_n$ and $p_n$ are the coordinate and momentum of bath oscillator $n$ with mass $m_n$ and frequency $\omega_n$. 
It is assumed that each bath oscillator is initially at thermodynamic equilibrium independent of the spin \textit{i.e.} relative to
\begin{equation}
    \label{eq:bath-classical-2}
  H_{\hbox{bath}} = \sum_n \frac{1}{2m_n} p_n^2 + \frac{1}{2}m_n\omega_n^2 x_n^2  
\end{equation}
The bath oscillators can then be integrated out and the Feynman-Vernon influence functional is
\begin{equation}
    \label{eq:FV}
    i\Phi = \frac{i}{\hbar} S_i[\cdot] - \frac{1}{\hbar} S_r[\cdot]
\end{equation}
where 
\begin{eqnarray}
  \label{eq:S_i}
  S_i &=& \int^{t_f}\int^t  (S_z^f(t)-S_z^b(t))(S_z^f(s)+S_z^b(s))k_i \\
  \label{eq:S_r}
  S_r &=& \int^{t_f}\int^t  (S_z^f(t)-S_z^b(t))(S_z^f(s)-S_z^b(s))k_r
\end{eqnarray}
The kernels $k_i$ and $k_r$ depend on the bath spectral density 
\begin{equation}
      \label{eq:bath-DOS}
  J(\omega)= \pi\sum_n \frac{C_n^2}{m_n\omega_n} \delta(\omega-\omega_n)   
\end{equation}
as
\begin{eqnarray}
  \label{eq:k_i}
  k_i &=& \frac{1}{\pi}\int_0^{\infty} J(\omega) \sin\omega(t-s) \\
  \label{eq:k_r}
  k_R &=& \frac{1}{\pi}\int_0^{\infty} J(\omega) \cos\omega(t-s) \coth(\frac{\hbar\omega}{2k_BT})
\end{eqnarray}
where $T$ is the temperature of the bath. 
In (\ref{eq:S_i}) and (\ref{eq:S_r}) for the
the forward spin history $S_z^f=\frac{1}{2}\cos\theta^f$ and for the backward $S_z^b=\frac{1}{2}\cos\theta^b$.
It is a consequence of the form of the coupling in the spin-boson model that the Feynman-Vernon action
only depends on the polar angle $\theta$ and not on the azimuthal angle $\phi$.

Assuming that $J(\omega)$ behaves as $\eta \omega^s \omega_c^{-s-1}$
up to some large frequency $\Omega$ and decays quickly for larger frequencies~\cite{Leggett87}
$k_i$ and $k_r$ will both be proportional to $\eta$ (in units of $\omega_c$). The kernel $k_i$ will have support on a time interval
of width $\Omega^{-1}$ and the kernel $k_r$ will have support on a time interval of
width the larger of  $\Omega^{-1}$ and $\hbar/k_B T$. Both (\ref{eq:S_i}) and (\ref{eq:S_r}) 
can therefore be simply estimated as
\begin{equation}
      \label{eq:S_i-S_r-estimate}
S_i,S_r \sim  \eta (t_f-t_i)  
\end{equation}
where $t_f-t_i$ is the duration of the process. This estimate is used above in the main text.  

\section{Analysis of recovery and quantum error correction}
\label{app:recovery}
In the main text of the paper quantum error correction 
was not considered. 
The purpose of this appendix is to discuss to what extent
an analysis based on Feynman-Vernon can be extended to a system with a recovery map.

The general conditions for successful quantum error correction were formulated by Knill and Laflamme~\cite{KnillLaflamme97}. 
The starting point is a code space ${\cal C}$
of which the ground state of the Kitaev model, Section~\ref{sec:toric-code} in main text, is
an example. The code space is a subset of a larger Hilbert space ${\cal H}$ called the coding space, and a super-operator $\Phi$ acts on 
density matrices on ${\cal H}$.
Perfect quantum error correction in ${\cal C}$ under $\Phi$ is possible if there
exist another super-operator ${\cal R}$, called a recovery operator,
such that ${\cal R}\Phi$ acts as the identity
on all pure states $|\Psi\rangle\langle\Psi |, \Psi\in {\cal C}$.
The general form of ${\cal R}$ is that of a measurement of the component in ${\cal H}$ orthogonal to ${\cal C}$, followed by a unitary transformation.
Alternatively, if $\Phi$ is represented in the Kraus form $\rho\to\sum_k A_k\rho A_k^{\dagger}$, 
the condition can be formulated as conditions on dynamical operators $A_k$ acting on  ${\cal C}$ (\cite{KnillLaflamme97}, Theorem III.2). 
If some quantum dynamics $\Phi$ on some space  ${\cal H}$ admits quantum error correction therefore
reduces to the question if there exists a code space ${\cal C}$. In general 
this is not trivial to decide, see \textit{e.g.}~\cite{Majgier10} for general rank-2 super-operators
and~\cite{Lipka-Bartosik17} for general 2-qubit maps. 

For error models often considered for the Kitaev code \textit{e.g.} in~\cite{DennisKitaevLandahlPreskill2002} and~\cite{Terhal15}
the above translates as follows.
The coding space ${\cal H}$ is that of all the physical qubits, and $\Phi$ has a block structure where each block acts on the states of one physical qubit. 
Interaction with the environment has hence been assumed to lead to super-operator which
is local in space (physical qubits) and also local in time (no memory),
compare discussion in Section~\ref{sec:quantum-noise} in main text.
By measuring stabilizers it is grosso modo possible to decide which unitary map was
applied, and then correct for it by applying its inverse. Precision to this statement,
consequences and concrete implementations have been discussed in great detail in the literature~\cite{Kitaev97,DennisKitaevLandahlPreskill2002,Fowler12}.
Within Feynman-Vernon theory the effects of random superposition of unitary transformations 
can be be described as follows:
if the influence action from applying $\rho\to\rho' = V_a\rho V_a^{\dagger}$ is $\Phi_a$, 
and if this transformation is applied with probability $p_a$, then the total influence action
is $\frac{1}{i}\log\sum_a p_a e^{i \Phi_a}$~\cite{FeynmanVernon}.
The total variational distance from (\ref{eq:TVD-approx})
is then
\begin{equation}
  \label{eq:TVD-approx-3}
  \hbox{TVD} \approx \sum_{\mathbf{f}} | \sum_a p_a \langle \Phi_a\rangle_{if} |
\end{equation}
If there are just a few unitary maps applied we are back to the same estimates
as in Section~\ref{sec:open-quantum-model}, but if there are many and they 
contribute with random phases the resulting TVD could be smaller due to cancellations.

Making additional assumptions we can also discuss the recovery map in the Kitaev code with error correction in the Feynman-Vernon formalism.
First, we assume that the stabilizers are measured very often but not absolutely
continuously. This is in line with proposed hardware implementations 
based on a system clock~\cite{Fowler12}, and implies that the degrees of the
freedom of both the logical qubits and the stabilizers can change between 
measurements. 
Second, the record of all the measurements of all the stabilizers is
assumed known. The corresponding chain of projection operators
acting on the full density matrix of the logical qubits, the stabilizers
and the environment is then a coarse-grained history in the sense of Gell-Mann and Hartle~\cite{Gell-MannHartle1993}.
Third, these coarse-grained histories are assumed 
to fulfill the decoherence conditions of~\cite{Gell-MannHartle1993}.
When (if) this is so we can consider the results of the measurements as 
known classical time-dependent parameters and write the Feynman-Vernon path integral
for the logical qubits and the environment as in Section~\ref{sec:toric-code}.
The difference for the spin-boson terms
would then be that while at most instance of time the appropriate form is
(\ref{eq:Q}) (when the measured values of the stabilizers do not change),
but sometimes it is (\ref{eq:Q-model}) (when they do). 

\bibliographystyle{spmpsci}      
\bibliography{fluctuations,quantum-computing}

\end{document}